\documentstyle[11pt,Moriond,epsfig,floatflt]{article}
\def\be{\begin{equation}}
\def\ee{\end{equation}}
\def\bea{\begin{eqnarray}}
\def\eea{\end{eqnarray}}

\def\vecmet{\mbox{$\vec{\not\!\!{E}}_T$}}

\def\ttb{\mbox{$t\bar{t}$}}

\def\vereq#1#2{\lower3pt\vbox{\baselineskip1.5pt \lineskip1.5pt}}

\def\beq{\begin{equation}}
\newcommand{\Ht}{\mbox{$H_{T}$}}
\newcommand{\et}{\mbox{$E_{T}$}}
\newcommand{\pt}{\mbox{$P_{T}$}}

\newcommand{\bspace}{\!\!\!\!}
\newcommand{\met}{\mbox{$E{\bspace}/_{T}$}}
\newcommand{\niso}{\mbox{$N_{trk}^{iso}$}}

\newcommand{\ppbar}{\mbox{$p\overline{p}$}}

\newcommand{\gev}{\mbox{GeV/$c^2$}}

\def\stilde{\widetilde}
\newcommand{\sq}{\stilde{q}}

\newcommand{\gls}{\stilde{g}}
\newcommand{\msq}{\mbox {$m_{\stilde{q}}$} }
\newcommand{\mgls}{\mbox {$m_{\tilde{g}}$} }

\def\eeq{\end{equation}}
\def\bea{\begin{eqnarray}}
\def\beaa{\begin{eqnarray*}}
\def\eea{\end{eqnarray}}
\def\eeaa{\end{eqnarray*}}
\def\bq{\begin{quote}}
\def\eq{\end{quote}}

\def\gappeq{\mathrel{\rlap {\raise.5ex\hbox{$>$}}
{\lower.5ex\hbox{$\sim$}}}}
\def\lappeq{\mathrel{\rlap{\raise.5ex\hbox{$<$}}
{\lower.5ex\hbox{$\sim$}}}}
\def\sm{Standard Model}
\begin{document}
\vspace*{0.8cm}
\begin{flushright}
\vspace*{-1.0in}
\hspace*{1.25in} 
\mbox EFI-01-23,FERMILAB-Conf-01/110-E\\ %
\end{flushright}
\title{SUSY SEARCHES AT THE TEVATRON}

\author{ M. SPIROPULU }

\address{Enrico Fermi Institute, 5640 S. Ellis Ave.,\\
Chicago IL, USA}

\maketitle\abstracts{
We discuss some of the latest results from supersymmetry searches at the Tevatron.
}
\section{Overview}
 D\O\ and CDF have already started collecting data from \ppbar\ collisions
at the Tevatron with the Main Injector in operation and at $\sqrt{s}\approx 2$ TeV. A number of recently completed SUSY searches using data from Run I are discussed.
\section{Search for scalar top production in electron plus muon plus \met\ channel at D\O\ }
\begin{floatingfigure} {2.7in}
\begin{center}
\epsfig{figure=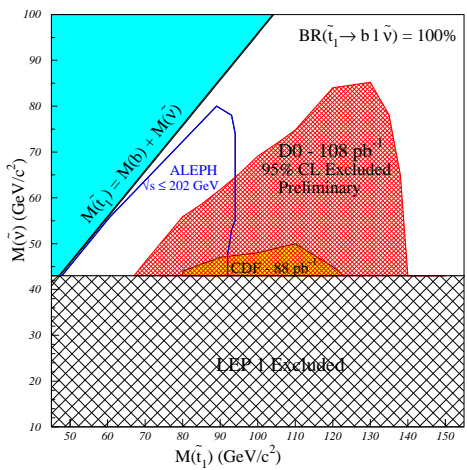,width=2.7in}
\end{center}
\label{gregorio}
\caption{Excluded region for $M(\tilde{t_1})$ in the $M(\tilde\nu)$
versus $M(\tilde{t_1})$ plane from $\tilde{t_{1}}\rightarrow b e^{\pm}(\mu^{\pm})\tilde{\nu}$ at D\O.}
\end{floatingfigure}
This is a search for direct  $\tilde{t_{1}}$$\bar{\tilde{t}}_{1}$
production where $\tilde{t_{1}}\rightarrow b \ell^{\pm}\tilde{\nu}$.
The branching ratio for each of the decays $\tilde{t_{1}}\rightarrow b e^{\pm}\tilde{\nu}$,
$\tilde{t_{1}}\rightarrow b \mu^{\pm}\tilde{\nu}$ and 
$\tilde{t_{1}}\rightarrow b \tau^{\pm}\tilde{\nu}$ is 33.3\%. The final state 
used in the search is an electron, a muon and missing transverse energy from
the $\tilde{\nu}$'s. 

The analysis uses 108$\pm$6 pb$^{-1}$ of data collected with the
D\O\ detector. The electron and muon are required to have \et$>$15 GeV, 
$|\eta|<2.5,1.7$ respectively and be acoplanar. The missing energy is 
required to be \met$>$15 GeV. The total number of expected \sm\ background events is 
13.4$\pm$1.5 mostly from fakes, $Z\rightarrow \tau\tau$, $t\bar t$ and dibosons. The number of
observed events in the data is 11. The analysis sets at 95\% C.L. upper limit 
on the stop pair production cross section as a function of the stop mass. In the $M(\tilde\nu)$
versus $M(\tilde{t_1})$ plane the bound is shown in Figure 1. Stop masses below 140 GeV/$c^2$ are excluded for sneutrino mass of 43 GeV/$c^2$ and below 130 GeV/$c^2$
for sneutrino mass of 85 GeV/$c^2$.

\section{Search for $R$-parity violating decays of the LSP in the dimuon plus
 4 jets final state at D\O}
The analysis uses 78 pb$^{-1}$ of data collected with the D\O\ detector and complements the corresponding search of $R$-parity violating decays of the LSP in the dielectron plus 4 jets final state\cite{Rd0}.
\begin{floatingfigure}{2.3in}
\vspace*{-0.9cm}
\begin{center}
\epsfig{figure=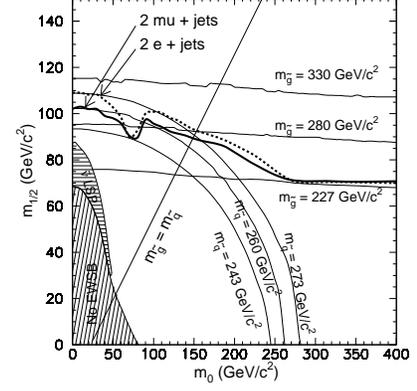,width=2.3in}
\end{center}
\label{sudeshna}
\caption{Exclusion curve in $m_{1/2}$ {\it vs} $m_{0}$ plane from
LSP $R$-parity violating decays.}
\end{floatingfigure}

Each of the LSPs of the final state decays 
via a lepton violating process into a lepton and two jets
($\tilde \chi_{1}^{0} \rightarrow e(\mu)qq^\prime$).
Two muons are required  with \pt$>$15, 10 GeV/$c$ respectively and  four jets with \et$>$15 GeV.
The total standard model background is estimated to be 0.18$\pm$0.03$\pm$0.02 (mostly 
from $Z\rightarrow \mu\mu$+jets) events. There are no events passing the selection criteria 
in the data. 

In the context of mSUGRA (with $A_0$=0, $\mu<$0, $\tan\beta=2$) the resulting bound is given in the common 
gaugino mass ($m_{1/2}$) versus common scalar mass  ($m_{0}$) plane as shown in Figure 2.

\section{Search for $R$-parity violating decays of scalar top quarks 
in di-$\tau$ plus dijet events at CDF}

This analysis searches for the scalar top  quark pair production
$\tilde{t_{1}}\overline{\tilde{t}}_{1}$ in which the stop decays via 
an $R$-parity violating process
into $\tau^{+}\overline b$. The search results are based on the 
final state of one electron from the one $\tau$ decay, 
one hadronically decaying $\tau$ ($\tau_{h}$) and two jets.
The final state is the same as in the 3rd generation leptoquark analysis
\cite{baumann}.
\begin{floatingfigure}{2.4in}
\vspace*{-0.2cm}
\begin{center}
\epsfig{figure=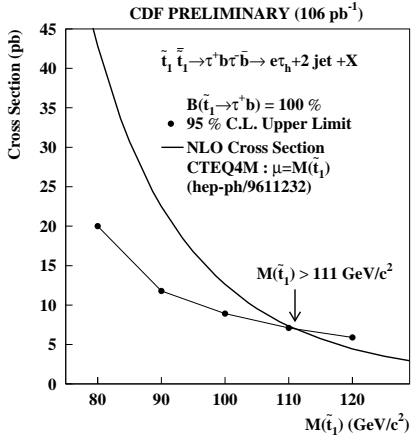,width=2.4in}
\end{center}
\label{miyazaki}
\caption{ The upper limit on the 
$\tilde{t_{1}}\overline{\tilde{t}}_{1}$  
cross section at 95\%
C.L. from $\tilde{t_{1}}\overline{\tilde{t}}_{1} \rightarrow \tau^{+} b\tau^{-} \bar{b}$
and the theory calculation as a function of the $\tilde{t_{1}}$ mass.}
\end{floatingfigure}
Using  a low \pt\ electron threshold (10 GeV) 
and a track based $\tau_{h}$ identification algorithm which includes
$\pi^{0}$ reconstruction, this analysis improves significally the $Z\rightarrow \tau\tau$ 
acceptance. After the baseline selection, the dominant sources
of $\ell+\tau_{h}$ events are found to be $Z\rightarrow\tau\tau$, $W$+jets, and 
QCD events. For the selection of  the  $\tilde{t_{1}}\overline{\tilde{t}}_{1}$
signal events and to eliminate the above backgrounds 
it is  required that the $(e,\met)$ transverse mass
be $\le 35$ GeV, the  scalar sum $\Ht \equiv \et(e)+\met+\pt(\tau_{h})\ge75$ GeV
and that the number of jets in the event is $N_{jet}\ge 2$.
To avoid systematics mainly due to the modeling of the $\tau$ identification,
the  $\tilde{t_{1}}\overline{\tilde{t}}_{1}$  production cross section is measured relative to the $Z\rightarrow\tau\tau$ cross section 
using the corresponding data sample. 
No events survive the selection while the standard model expectation
is 1.91$\pm$0.11$\pm$0.15 mainly from $Z\rightarrow\tau\tau$.
The analysis excludes at 95\% C.L. $\tilde{t_1}$ masses below 111 GeV/$c^2$ as shown in
Figure 3.

\section{A model independent search for new physics with leptons and photons
in the final state at CDF}

The unexplained $ee\gamma\gamma\met$ event observed  at CDF, pointed
towards  possible  novel processes involving combinations 
of leptons and photons in the final state.

This analysis searches for anomalous production of events with a high \et\
photon and a lepton ($e$ or $\mu$) in the final state. 

The search is based on \mbox{86 pb$^{-1}$} of
data collected with the CDF detector during the 1994-95 Tevatron run.
The cases with  large \met, additional photons or additional leptons
in the final state are also analyzed.

\begin{floatingtable}{
\begin{tabular}{lrrc}\hline\hline
Category & 
\multicolumn{1}{c}{$\mu_{SM}$} & 
\multicolumn{1}{c}{$N_0$} & 
\multicolumn{1}{c}{$P(N\geq N_0|\mu_{SM})$ \%} \\ \hline
All $\ell\gamma X$& \multicolumn{1}{c}{---}  & 77 & --- \\
\hline
Z-like $e\gamma$ & \multicolumn{1}{c}{---}  & 17 & --- \\
Two-Body $\ell\gamma X$        & $24.9\,\ \pm2.4\,\ $ & 33 &  $\,\  9.3$ \\ 
Multi-Body $\ell\gamma X$      & $20.2\,\ \pm1.7\,\ $ & 27 & 10.0 \\ 
\hline
Multi-Body $\ell\ell\gamma X$  & $5.8\,\ \pm0.6\,\ $   &  5 & 68.0 \\ 
Multi-Body $\ell\gamma\gamma X$& $0.02\pm0.02$ &  1 & $\,\  1.5$ \\ 
Multi-Body $\ell\gamma\met X$ & $7.6\,\ \pm0.7\,\ $   & 16 & $\,\  0.7$ \\ 
\hline\hline
\end{tabular}}
\caption{The results for all photon-lepton categories analyzed, including 
the mean number of events $\mu_{SM}$ predicted by the standard model,
the number $N_0$ observed in CDF data, 
and the observation likelihood $P(N\geq N_{0}|\mu_{SM})$.} 
\label{lgamma_result}
\end{floatingtable}
\noindent A two-body photon-lepton, a multi-body photon-lepton sample, 
and several subsets of the multi-body photon-lepton 
sample with additional particles were studied and the results
were consistent with standard model expectations with a possible 
exception in a subset of the multi-body photon-
lepton sample, consisting
of those events with $\met > 25$ GeV. The results of the 
analysis of these samples are shown in Table ~\ref{lgamma_result}\cite{JeffB}.

\section{Search for Gluinos and Scalar Quarks at CDF using the Missing Energy plus
Multijets Signature}
\begin{floatingfigure}{3.2in}
\begin{center}
\epsfig{figure=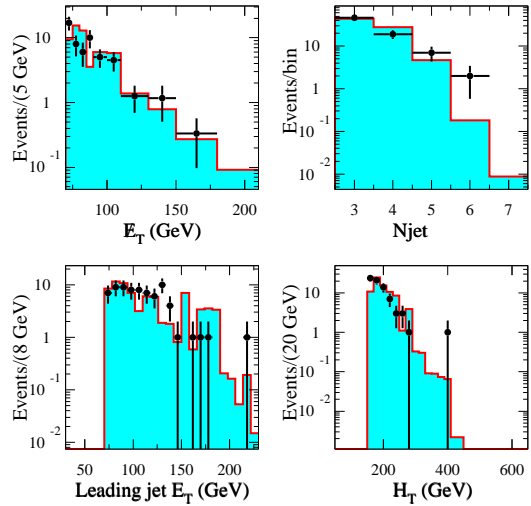,width=3.2in}
\end{center}
\label{smaria1}
\caption{ Comparison in the `blind box' 
between data (points) and \sm\ predictions (histogram) of ~\met,
$N_{jet}$, leading jet \et\ and \Ht\ distributions. There are 74
events in each of these plots, to be compared with 76$\pm$13 SM
predicted events.}
\end{floatingfigure}
This analysis  investigates whether the production and decay of gluinos and 
scalar quarks  is observable in the rate of ~$\ge$3-jet events
with large missing transverse energy  at the CDF.   
The large missing energy would originate from the two LSPs 
in the final states of the squark and gluino 
decays. The three or more  hadronic jets  would result 
from the hadronic decays of the $\sq$ and/or $\gls$.
The search is based on \mbox{$84 \pm 4$ pb$^{-1}$} of
integrated luminosity recorded with the CDF detector
during the 1994-95 Tevatron run.

The data sample was selected with an on-line trigger which 
requires $\met \equiv |\vecmet| > 30$ GeV.
A two-stage preselection  rejects accele-\\rator-, detector-related backgrounds, beam halo, and cosmic ray events\cite{smaria_paper}.
At least three jets with $E_T\ge 15$ GeV, at least one of them within
$|\eta|< 1.1$, are then required in events that pass the
preselection. In a QCD multijet event with large missing energy, 
the highest \et\ jet is typically the most 
accurately measured.  When the second 
or third jet is mismeasured because it lands 
partially in  an uninstrumented region (a `gap'),
the \met\ is pulled close in $\phi$ to the mismeasured 
jet. A jet is considered non-fiducial if it is within 0.5 rad in
$\phi$ of the \met\ direction and also points in $\eta$ to a detector
gap.  In this analysis the second and third highest \et\ jets in an event are
required to be fiducial.  The residual QCD component is eliminated by
using the correlation in the $\delta\phi_{1}=|\phi_{{\rm leading
~jet}}-\phi_{\met}|$ versus $\delta\phi_{2}= |\phi_{{\rm second~
jet}}-\phi_{\met}|$ plane and  by requiring the \met\  not be closer than 0.3 rad in
$\phi$ to any jet in the event. The background contribution from $W(\rightarrow e\nu)+$jets
and \ttb\ production is reduced by requiring the two highest energy jets not be
purely electromagnetic (jet electromagnetic fraction $f_{em}<$0.9).
The two highest \et\ jets in the event are required to be  $E_{T(1)} \ge$70 GeV and
$E_{T(2)} \ge$30 GeV.
\begin{floatingfigure}{3.2in}
\label{smaria2}
\begin{center}
\epsfig{figure=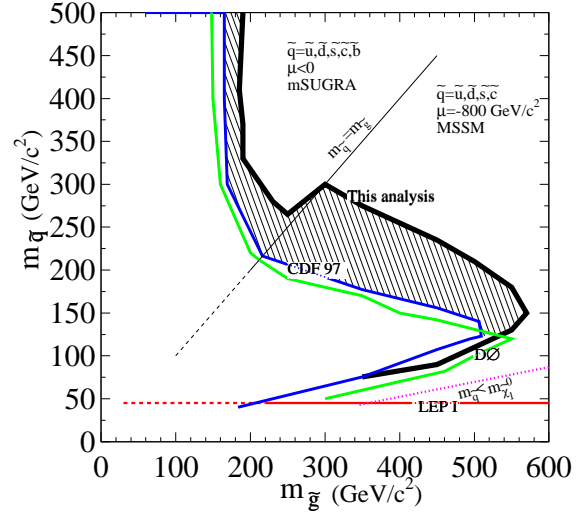,width=3.1in}
\end{center}
\caption{The 95\% C.L. limit curve in the $\msq~-~\mgls$ plane 
for $\tan\beta=3$; the hatched area is newly excluded by this
analysis. Results from some previous searches are also
shown. The region labelled as $\msq < m_{\stilde{\chi}_{1}^{0}}$ is
theoretically forbidden as the squarks are predicted to be lighter
than the LSP.}
\end{floatingfigure}
To avoid potential {\it a posteriori} biases when searching for new
physics in the tails of the missing transverse energy distribution, this
analysis defines and seals the signal candidate data sample. This
analysis approach is often referred to as a `blind analysis' and the
signal candidate data sample as a `blind box'. The `blind box' data
are inspected only after the entire search path has been defined by
estimating the total \sm\ backgrounds and
optimizing the sensitivity to the supersymmetric
signal. Three variables define the signal candidate region :
\met, $\Ht \equiv E_{T(2)}+E_{T(3)}+\met$, and isolated track
multiplicity, \niso ~\cite{smaria_paper}.  The `blind box' contains events
with $\met\ge 70$ GeV, \mbox{$H_T\ge 150$ GeV}, and \niso=0.  
Events with large missing transverse energy and
$\ge$3  jets in the final state are expected primarily from
QCD, $Z(\rightarrow \nu\bar{\nu})+$\\$\ge$3 jets, $W(\rightarrow
\tau\nu)+\ge$2 jets (the third jet originating from the hadronic
$\tau$ decay) and $t\bar t$ processes. 
Of the 76 events predicted in the `blind box', 41 come from QCD 
and 35 from electroweak processes. Of the latter $\sim$37\% are expected from 
$Z\rightarrow \nu\bar{\nu}+\ge 3$ jets, $\sim$20\% from
$W\rightarrow \tau\nu+\ge 2$ jets, $\sim$20\% from the combined
$W\rightarrow e(\mu)\nu_{e}(\nu_{\mu})+\ge 3$ jets, and $\sim$20\%
from \ttb\ production and decays. The \met\ and \Ht\ requirements are optimized to
increase sensitivity to the signal in the context of MSSM and mSUGRA models and as a function of  $\msq\over\mgls$. 

In the `blind box', where 76$\pm$13 \sm\ events are expected,
74 events are observed.  In Figure 4 the predicted \sm\ kinematic
distributions are compared with the distributions observed in the
data. The derived 95\% C.L. bound is shown on the $\msq-\mgls$ plane in
Figure 5. The search excludes gluino masses below \mbox{195 \gev}
(95\% C.L.), independent of the squark mass.  For the case $\msq
\approx \mgls$, gluino masses below \mbox{300 \gev} are excluded.
\section*{Acknowledgments}
Many thanks to the organizers, C. Barthelemy and the {\em Improving Research Potential and the Socio-economic Knowledge Base; High Level Scientific conferences; HPCF-CT-1999-0028} European grant for partial support towards attendanting the conference. Also G. Brooijmans, R. Culbertson,  G. Bernardi, S.
Banerjee, Y. Miyazaki and J. Berryhill.


\section*{References}
\begin{thebibliography}{99}
\bibitem{Rd0}B.Abbott {\it et al.}, The D0 Collaboration Phys. Rev. Lett. {\bf 83}, 4476 (1999). 
\bibitem{baumann} F. Abe {\it et al.}, The CDF Collaboration, Phys. Rev. Lett. {\bf 78}, 2906 (1997).
\bibitem{JeffB} J.Berryhill http://moriond.in2p3.fr/EW/2001/proceedings/
\bibitem{smaria_paper} T. Affolder {\it et al.}, The CDF Collaboration, hep-ex/0106001 (2001).
\end{thebibliography}
\end{document}